\documentclass[preliminary,
]{eptcs}
\pdfoutput=1

\usepackage[dvips]{epsfig}
\usepackage{color}
\usepackage{amsmath}
\usepackage{amssymb}
\usepackage[dvips]{graphicx}
\usepackage{stmaryrd}
\usepackage{bussproofs}
\newcommand{\parsym}{\bindnasrepma}

\newtheorem{definition}{\bf Definition}
\newtheorem{theorem}{\bf Theorem}
\newtheorem{proposition}{\bf Proposition}
\newtheorem{lemma}{\bf Lemma}

\newtheorem{example}{\bf Example}

\title{Bipolar Proof Nets for MALL\thanks{Research supported by the PRIN Project {\sc Concerto}.}\\
{\small\em Proceedings of the PCC12 Conference -- 17-18 August 2012, University of Copenhagen, Denmark}}
\author{
Roberto Maieli
\institute{Universit\`a degli Studi ``Roma Tre''}
\email{maieli@uniroma3.it}}

\begin{document}
\maketitle

\begin{abstract}
In this work we present a computation paradigm based on a concurrent and  incremental construction of proof nets (de-sequentialized or graphical proofs) of the pure multiplicative and additive fragment of Linear Logic, a resources conscious refinement of Classical Logic. Moreover, we set a correspondence between this paradigm and those more pragmatic ones inspired to transactional or distributed systems. In particular we show that the construction of additive proof nets can be interpreted as a model for {\em super-ACID} (or {\em co-operative}) transactions over distributed 
transactional systems (typically, multi-databases).
\end{abstract}

\section{Introduction}
%%%%%%%%%%%%%%%%%%%%%%
\label{sec:intro}

This work takes a further step towards the development of an ambitious research programme, firstly started by Andreoli in~\cite{jma01}, 
which aims at a theoretical foundation of a {\em computational programming paradigm based on 
the construction of proofs of linear logic} (LL,~\cite{gir87}). Naively, this paradigm relies on the following isomorphism: 
``proof''=``state'' and ``construction step (or inference)''=``state transition''. 

While the view of proof construction is well adapted to theorem proving, it is inadequate when we want to model the execution of 
widely distributed applications (typically over the Internet) which are designed with very flexible, concurrent and modular approaches.
Due to their artificial sequential nature,  sequent proofs are difficult to cut into composable (reusable) concurrent modules.
A much more appealing solution consists in using the technology offered by {\em proof nets} of linear logic or, more precisely, some forms of de-sequentialized (geometrical indeed) proof structures 
in which the composition operation is simply given by (constrained) juxtaposition, obeying to some correctness criteria. 

%% Here we show how much powerful this paradigm can become when it relies on the key concept of linear logic: proof nets, that is, de-sequantialized or grafical proofs. 

Actually, the proof net construction, as well as  the proof net cut reduction, 
can be performed in parallel (concurrently), but despite from the cut reduction, there may not exist executable (sequentializable) 
construction steps: in other words, construction steps must satisfy an ``efficient'' correction criterion. The resulting paradigm is very close to more
pragmatic ones, like those ones coming from transactional or distributed systems.

Concretely, here, we present a model for the incremental construction of proof nets of the {\em pure multiplicative and additive fragment of linear logic} (MALL,~\cite{gir96}). 
This model extends the previous one, given in~\cite{jma02}, for the {\em pure multiplicative fragment of linear logic} (MLL). In particular,
we give a syntax for {\em bipolar focussing  proof-structures} that are de-sequentialized (geometrical) representations of 
possibly {\em incomplete} ({\em open} or {\em with proper axioms}) proofs of the {\em bipolar focussing sequent calculus}~\cite{jma01}. 
This calculus has the following properties:
\begin{enumerate}
 \item\label{1} the possibly incomplete (open) focussing proofs are strictly isomorphic to the possibly open proofs of the 
bipolar focussing sequent calculus;
 \item\label{2} the complete (closed or with logical axioms) focussing proofs are fully representative of all the closed proofs of linear logic.
\end{enumerate}
Hence by~\ref{1} and~\ref{2}, proof construction can be performed equivalently in these three proof systems of LL: {\em sequent calculus}, {\em focussing sequent calculus} and
       {\em bipolar focussing sequent calculus}. Bipolarity and focussing properties ensure 
more compact proofs since they get rid of some irrelevant intermediate steps in the construction.

In~\cite{jma02,jma03}, the concurrent construction of open (transitory) MLL proof nets 
was interpreted  as an incremental juxtaposition of link modules (agents) that allows to model the behavior of {\em ACID transactions} over strongly distributed systems.
Here the proof construction of transitory MALL proof nets is interpreted as an additive  (super) juxtaposition of interacting 
{\em slices} (multiplicative transitory proof nets). Locally the concurrent construction of MALL proof nets can be viewed as an incremental 
juxtaposition of  hyperlinks (a disjoint sum of multiplicative links) that, like co-operative agents,  allow to model some kinds of 
(non-deterministic) co-operation among ACID transactions.

\section{Bipolar Focussing Sequent Calculus}
%%%%%%%%%%%%%%%%%%%%%%%%%%%%%%%%%%
\label{sec:bsq}

We recall some basic definitions of the {\em standard sequent calculus of MALL}, then we introduce the related {\em bipolar focussing  sequent calculus}, based on the crucial properties of 
{\em focussing} and {\em bipolarity} (find more in~\cite{jma01},~\cite{gir01} and~\cite{lau99}).
We, arbitrarily assume literals $a,a^\perp,b,b^\perp,...$ with a polarity: {\em negative} for atoms and {\em positive} for their duals, then 
given a set $\cal A$ of atoms, an $\cal A$-formula is a formula built from  atoms and their duals, using the (two groups of) connectives of MALL:
 {\em negative}, $\parsym$ ("par") and $\&$ ("with") and {\em positive}, $\otimes$ ("tensor") and $\oplus$ ("plus"). Finally, 
%\begin{description}
% \item{\em negative} - $\parsym$ ("par") and $\&$ ("with");
% \item{\em positive} - $\otimes$ ("tensor") and $\oplus$ ("plus").
%\end{description}
a proof of MALL is build by means of the following (groups of) inferences:

\begin{center}
{\em identity} : \;\;
% \resizebox{0.52\textwidth}{!}{
\AxiomC{}
\RightLabel{ax}
\UnaryInfC{$A,A^\perp$}
\DisplayProof
\hspace{2cm}
{\em multiplicatives} : \;\;
% \resizebox{0.50\textwidth}{!}{
\AxiomC{$\Gamma,A$}
\AxiomC{$\Delta,B$}
\RightLabel{$\otimes$}
\BinaryInfC{$\Gamma,\Delta,A\otimes B$}
\DisplayProof
\;\;\;
\AxiomC{$\Gamma,A,B$}
\RightLabel{$\parsym$}
\UnaryInfC{$\Gamma,A\parsym B$}
\DisplayProof

\medskip
{\em  additives} : \;\;
% \resizebox{0.82\textwidth}{!}{
\AxiomC{$\Gamma,A$}
\AxiomC{$\Gamma,B$}
\RightLabel{$\binampersand$}
\BinaryInfC{$\Gamma,A\binampersand B$}
\DisplayProof
\;\;\;
\AxiomC{$\Gamma,A$}
\RightLabel{$\oplus_1$}
\UnaryInfC{$\Gamma,A\oplus_1 B$}
\DisplayProof
\;\;\;
\AxiomC{$\Gamma,B$}
\RightLabel{$\oplus_2$}
\UnaryInfC{$\Gamma,A\oplus_2 B$}
\DisplayProof
\end{center}

%\medskip
%\begin{tabular}{lccc}
%%% \begin{center}
%{\em identity} : &
%% \resizebox{0.52\textwidth}{!}{
%\AxiomC{}
%\RightLabel{ax}
%\UnaryInfC{$A,A^\perp$}
%\DisplayProof

%& 

%\AxiomC{$\Gamma,A$}
%\AxiomC{$\Delta,A^\perp$}
%\RightLabel{cut}
%\BinaryInfC{$\Gamma,\Delta$}
%\DisplayProof
%%}

%\\
%\\

%% \vspace{1cm}
%{\em multiplicatives} : &
%% \resizebox{0.50\textwidth}{!}{
%\AxiomC{$\Gamma,A$}
%\AxiomC{$\Delta,B$}
%\RightLabel{$\otimes$}
%\BinaryInfC{$\Gamma,\Delta,A\otimes B$}
%\DisplayProof

%&

%\AxiomC{$\Gamma,A,B$}
%\RightLabel{$\parsym$}
%\UnaryInfC{$\Gamma,A\parsym B$}
%\DisplayProof
%% }

%\\
%\\

%% \vspace{1cm}
%{\em  additives} : &
%% \resizebox{0.82\textwidth}{!}{
%\AxiomC{$\Gamma,A$}
%\AxiomC{$\Gamma,B$}
%\RightLabel{$\binampersand$}
%\BinaryInfC{$\Gamma,A\binampersand B$}
%\DisplayProof

%& 

%\AxiomC{$\Gamma,A$}
%\RightLabel{$\oplus_1$}
%\UnaryInfC{$\Gamma,A\oplus_1 B$}
%\DisplayProof

%& 

%\AxiomC{$\Gamma,B$}
%\RightLabel{$\oplus_2$}
%\UnaryInfC{$\Gamma,A\oplus_2 B$}
%\DisplayProof
%% }
%%% \end{center}
%\end{tabular}

\medskip
The {\em focussing property} states that, in the proof search (or proof construction), we can build (bottom up) a sequent proof by alternating clusters of negative inferences followed by clusters of positive inferences. As consequence of this bipolar alternation we obtain more compact proofs in which we get rid of the most part of all the bureaucracy hidden in sequential proofs (as, for instance, irrelevant permutation of rules): what remains is a focussing bipolar proof. Remind that w.r.t. proof search  negative (resp., positive) connectives involve  a kind of {\em don't care non-determinism} 
 (resp., {\em true non-determinism}).

\smallskip
An $\cal A$-{\em monopole} is an $\cal A$-formula built on negative $\cal A$-atoms using only the negative connectives; an $\cal A$-{\em bipole} is 
an $\cal A$-formula built from $\cal A$-monopoles and positive $\cal A$-atoms, using only positive connectives; moreover, bipoles must contain at least 
one positive connective or be reduced to a positive atom, so that they are always disjoint from monopoles.

Given a set $\cal F$ of $\cal A$-bipoles, the {\em bipolar focussing sequent calculus} $\Sigma[\cal A,\cal F]$ is a set of inferences of the form:
\begin{prooftree} 
\AxiomC{$\Gamma_1$}
\AxiomC{$\dots$}
\AxiomC{$\Gamma_n$}
\RightLabel{$F$}
\TrinaryInfC{$\Gamma$}
\end{prooftree}
where the conclusion $\Gamma$ is a sequent made by only of negative $\cal A$-atoms and the premises $\Gamma_1,...,\Gamma_n$ are obtained by fully 
focussing decomposition of some bipole $F\in{\cal F}$ in the the context $\Gamma$ (a multiset of negative atoms). 
More precisely, due to the presence of additives (in particular the $\oplus$ connectives) 
a bipole $F$ is naturally associated to a set of inferences $F_1,...,F_{m+1}$, where $m$ is the number of $\oplus$ connectives presents in $F$. 
For instance, in the purely multiplicative fragment of LL, the bipole $F = a^\perp\otimes b^\perp\otimes (c\parsym d)\otimes e$, 
where $a,b,c,d,e$ are (negative) $\cal A$-atoms, yields the inference on the left-hand side (more compact w.r.t. the explicit one on the right hand side):
\begin{prooftree} 
\AxiomC{$\Gamma,c,d$}
\AxiomC{$\Delta,e$}
\RightLabel{$F$}
\BinaryInfC{$\Gamma,\Delta,a,b$}
\DisplayProof
\;\;\;\;\;\;$\Leftrightarrow$\;\;\;\;\;\;
\AxiomC{$\Gamma,c,d$}
\RightLabel{$\parsym$}
\UnaryInfC{$\Gamma,c\parsym d$}
\AxiomC{$\Delta,e$}
\RightLabel{$\otimes$}
\BinaryInfC{$\Gamma,\Delta,(c\parsym d)\otimes e$}
\AxiomC{$b,b^\perp$}
\AxiomC{$a,a^\perp$}
\RightLabel{$\otimes$}
\TrinaryInfC{$\Gamma,\Delta,a, b,{\bf a^\perp\otimes b^\perp\otimes(c\parsym d)\otimes e}$}
\end{prooftree}
where $\Gamma,\Delta$ rage over a multiset of negative $\cal A$-atoms. Note that the identity axioms $a,a^\perp$ and $b,b^\perp$ are omitted 
in the bipolar sequent proof for simplicity sake. The couple  $a,b$ here plays the role of a {\em trigger} or {\em mutlifocus} of the $F$-inference; more generally, 
a trigger of a bipole is a multiset of duals of the positive atoms which occurs in it. The main feature of the bipolar focussing  sequent calculus is 
that its inferences are triggered by multiple focus (like in Forum~\cite{miller96}).

The bipolar focussing  sequent calculus is proved (Theorem~\ref{th:univprog}, see~\cite{jma01}) to be isomorphic to the focussing sequent calculus, 
so that proof construction can be performed indifferently  in the two systems.  
The main idea exploited in the proof of Theorem~\ref{th:univprog} is the {\em bipolarisation technique}, that is a simple procedure that allows
to transform a provable formula $F$ in the LL sequent calculus into a set of bipoles (belonging to an ``universal program'' in the 
bipolar sequent calculus). For our purpose, we briefly illustrate this technique only for the MALL fragment, with an instance given in the Example\ref{example:2}.

\smallskip
A {\em naming scheme} is a triple $\langle{\cal A},{\cal A'},\eta\rangle$ where $\cal A\subset\cal A'$ are sets of negative atoms and $\eta$ is a bijection 
from the $\cal A$-formulas into $\cal A'$ such that $\eta_a=a$ for all $a\in\cal A$. The {\em universal program} for a naming scheme 
$\langle{\cal A},{\cal A'},\eta\rangle$ is the set of $\cal A'$-bipoles of the form $\nu(F)$ where $F$ ranges over the $\cal A$-formulas not 
reduced to a negative atom. The $\nu$-mapping on $\cal A$-formulas is defined in three steps as follows:
\begin{enumerate}
 \item ({\em negative layer}) mapping $\nu^\uparrow$ from $\cal A$-formulas to $\cal A'$-monopoles
\begin{center}
\begin{tabular}{rcll}
 $\nu^\uparrow(F_1\parsym F_2)$ & $=$ & $\nu^\uparrow(F_1)\parsym\nu^\uparrow(F_2)$\\
 $\nu^\uparrow(F_1\& F_2)$      & $=$ & $\nu^\uparrow(F_1)\&\nu^\uparrow(F_2)$\\
 $\nu^\uparrow(F)$              & $=$ & $\eta_F$ in all the other cases;
\end{tabular}
\end{center}
 \item ({\em positive layer}) mapping $\nu^\downarrow$ from $\cal A$-formulas to $\cal A'$-bipoles or monopoles
\begin{center}
\begin{tabular}{rcll}
 $\nu^\downarrow(F_1\otimes F_2)$ & $=$ & $\nu^\downarrow(F_1)\otimes\nu^\downarrow(F_2)$\\
 $\nu^\downarrow(F_1\oplus F_2)$  & $=$ & $\nu^\downarrow(F_1)\oplus\nu^\downarrow(F_2)$\\
 $\nu^\downarrow(a^\perp)$        & $=$ & $a^\perp$ if $a$ is a negative atom\\
 $\nu^\downarrow(F)$              & $=$ & $\nu^\uparrow(F)$ in all the other cases;
\end{tabular}
\end{center}
 \item mapping $\nu$ from $\cal A$-formulas to $\cal A'$-bipoles
\begin{center}
\begin{tabular}{rcll}
 $\nu(F)$ & $=$ & $\eta^\perp_F\otimes\nu^\downarrow(F)$.
\end{tabular}
\end{center}
\end{enumerate}

\begin{theorem}[universal program]~\label{th:univprog}
Given a naming scheme $\langle{\cal A},{\cal A'},\eta\rangle$, let $\cal U$ be its universal program.
For any $\cal A$-formula $F$ there is an isomorphism between the focussing proofs of $F$ in linear logic and 
the proofs of $\eta_F$ in the bipolar focussing sequent calculus $\Sigma[\cal A',U]$.
\end{theorem}

\begin{example}\label{example:2}
%%This example illustrates how does work the bipolarisation. 
Assume an $\cal A$-formula $F=(a\& b)\parsym ((a^\perp\oplus b^\perp)\otimes c^\perp)\parsym(c\otimes(d^\perp\oplus e^\perp))\parsym(d\& e)$ with 
subformulas $G=(a^\perp\oplus b^\perp)\otimes c^\perp$ and $H=c\otimes(d^\perp\oplus e^\perp)$ and negative atoms $a,b,c,d,e$. 
After bipolarisation of $F$ we get the following bipoles of the universal program $\cal U$:

\begin{center}
\begin{tabular}{rcll}
 $\nu(F)$ & $=$ & $\eta^\perp_F\otimes((a\& b)\parsym\eta_G\parsym\eta_H\parsym(d\& e))$ \\
 $\nu(G)$ & $=$ & $\eta^\perp_G\otimes((a^\perp\oplus b^\perp)\otimes c^\perp)$\\
 $\nu(H)$ & $=$ & $\eta^\perp_H\otimes c\otimes(d^\perp\oplus e^\perp)$;
\end{tabular}
\end{center}

-- the bipole $\nu(F)$ corresponds to  the unique inference $\nu(F)$:

\begin{center}
$\frac{
\AxiomC{}
\noLine
% \RightLabel{$\nu(G)$}
% \UnaryInfC{$\eta_G,a,c$}
% \RightLabel{$\nu(H)$}
\UnaryInfC{$\Gamma,\eta_G,\eta_H,a,d$}
\DisplayProof
\hspace{0.5cm}
\AxiomC{}
\noLine
% \RightLabel{$\nu(G)$}
% \UnaryInfC{$\eta_G,a,c$}
% \RightLabel{$\nu(H)$}
\UnaryInfC{$\Gamma,\eta_G,\eta_H,a,e$}
\DisplayProof
\hspace{0.5cm}
\AxiomC{}
\noLine
% \RightLabel{$\nu(G)$}
% \UnaryInfC{$\eta_G,b,c$}
% \RightLabel{$\nu(H)$}
\UnaryInfC{$\Gamma,\eta_G,\eta_H,b,d$}
\DisplayProof
\hspace{0.5cm}
\AxiomC{}
\noLine
% \RightLabel{$\nu(G)$}
% \UnaryInfC{$\eta_G,b,c$}
% \RightLabel{$\nu(H)$}
\UnaryInfC{$\eta_G,\eta_H,b,$e}
\DisplayProof}
{\AxiomC{}
\noLine
%% \RightLabel{$\nu_F$}
\UnaryInfC{$\Gamma,\eta_F$}
\DisplayProof
}{\nu(F)}$
\end{center}

 -- the bipole $\nu(G)$ is associated to a pair of inferences: %% $(\nu_G)_1$ and $(\nu_G)_2$ 

\begin{center}
\AxiomC{$\Gamma$}
\RightLabel{$\nu(G)_1$}
\UnaryInfC{$\Gamma,\eta_G,a,c$}
\DisplayProof
\hspace{0.5cm} 
and 
\hspace{0.5cm}
\AxiomC{$\Gamma$}
\RightLabel{$\nu(G)_2$}
\UnaryInfC{$\Gamma,\eta_G,b,c$}
\DisplayProof
\end{center}

 -- similarly, the bipole $\nu(H)$ is associated to a pair of inferences: %% $(\nu_H)_1$ and $(\nu_H)_2$

\begin{center}
\AxiomC{$\Gamma,c$}
\RightLabel{$\nu(H)_1$}
\UnaryInfC{$\Gamma,\eta_H,d$}
\DisplayProof
\hspace{0.5cm} 
and
 \hspace{0.5cm} 
\AxiomC{$\Gamma,c$}
\RightLabel{$\nu(H)_2$}
\UnaryInfC{$\Gamma,\eta_H,e$}
\DisplayProof
\end{center}

Finally, here is the complete bipolar focussing proof of $\eta_F$ that is isomorphic, by Theorem~\ref{th:univprog}, to the 
(omitted) proof of $F$ in the LL focussing sequent calculus:

\begin{center}
$\frac{
\AxiomC{}
\RightLabel{$\nu(G)_1$}
\UnaryInfC{$\eta_G,a,c$}
\RightLabel{$\nu(H)_1$}
\UnaryInfC{$\eta_G,\eta_H,a,d$}
\DisplayProof
\hspace{0.5cm}
\AxiomC{}
\RightLabel{$\nu(G)_1$}
\UnaryInfC{$\eta_G,a,c$}
\RightLabel{$\nu(H)_2$}
\UnaryInfC{$\eta_G,\eta_H,a,e$}
\DisplayProof
\hspace{0.5cm}
\AxiomC{}
\RightLabel{$\nu(G)_2$}
\UnaryInfC{$\eta_G,b,c$}
\RightLabel{$\nu(H)_1$}
\UnaryInfC{$\eta_G,\eta_H,b,d$}
\DisplayProof
\hspace{0.5cm}
\AxiomC{}
\RightLabel{$\nu(G)_2$}
\UnaryInfC{$\eta_G,b,c$}
\RightLabel{$\nu(H)_2$}
\UnaryInfC{$\eta_G,\eta_H,b,e$}
\DisplayProof}
{\AxiomC{}
\noLine
%% \RightLabel{$\nu_F$}
\UnaryInfC{$\eta_F$}
\DisplayProof
}{\nu(F)}$
\end{center}
\end{example}

Observe that while the above derivation is quite compact, it still presents a lot of structural inconvenient such as duplications of sub-trees; phenomena like these are
 crucial when we want to modelize the behavior of distributed systems. For these reasons,  in the next section, we move to more 
flexible (geometrical indeed) proof structures.

\section{Bipolar Focussing Proof Structures}
%%%%%%%%%%%%%%%%%%%%%%%%%%%%%%%%%%
\label{sec:bps}

In this section we introduce the sequentialized version of the bipolar focussing sequent calculus, i.e. a graphical representation 
of bipolar proofs as {\em proof-structures} (eventually correct, i.e. {\em proof nets}) which preserves only essential sequentializations.

\begin{definition}[links]\label{def:links}
Assume an infinite set $\cal L$ of {\em resource places} $l_1,l_2,...$ (also {\em addresses} or {\em loci} like in Ludics~\cite{gir01}); the special untyped place $\star$ is called {\em jump place}.
A {\em link} consists in two disjoint sets of loci, top and bottom, together with a polarity $p$, positive  or negative,  and 
with the conditions that:
\begin{itemize}
\item a positive link must have {\em at least one}  bottom place; it may contain no more than one jump place among its bottom places;
\item a negative link must have {\em exactly one}  bottom place; it may contain no more than one jump place among its top places.
\end{itemize}

If the set of top places is not empty, then a link is said {\em transitional}. %% Bottom and top places are labelled by occurrences of negative atoms.
\end{definition}

\begin{figure}%%%%%%%%%%%%%%%%%%%%%%%%%%%%%%%%%%%%%%%%%%%%%%%
\begin{center}
\resizebox{0.25\textwidth}{!}{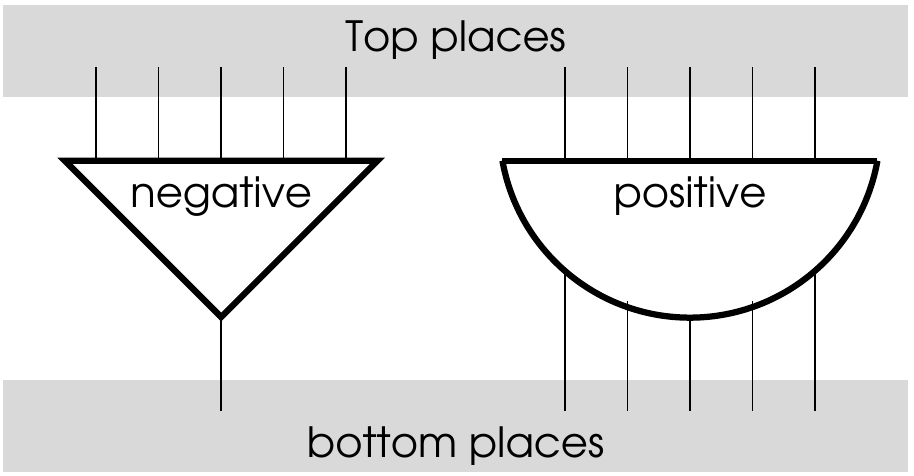} \hspace{1cm}                                          %%% links
\resizebox{0.16\textwidth}{!}{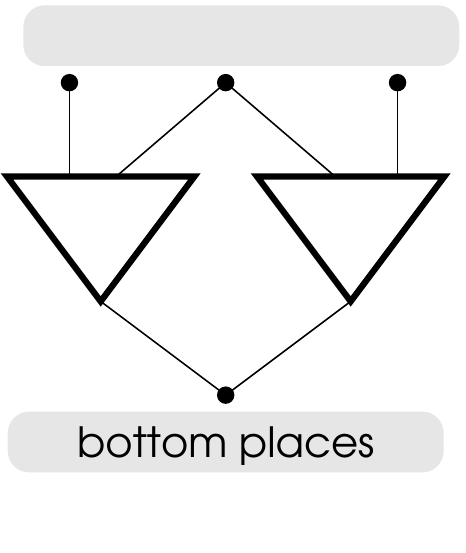}\hspace{2cm}                                 %%% negative distributivity
\resizebox{0.20\textwidth}{!}{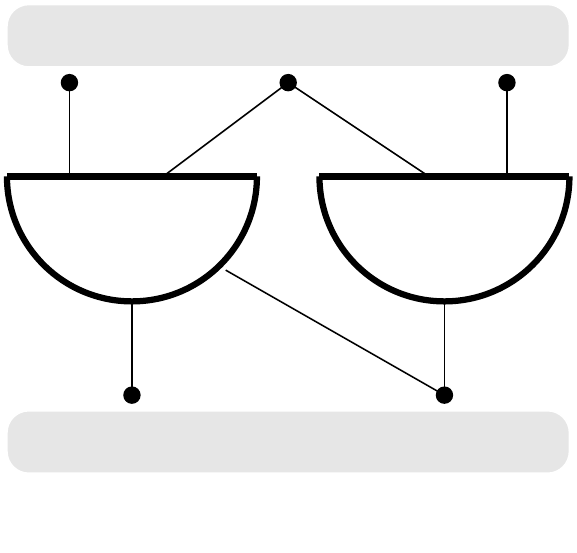}                                                         %%% positive distributivity
%\framebox{\resizebox{0.20\textwidth}{!}{\input{pics/bipolar00.pstex_t}}}\hspace{1cm}       %%% bipole
%\framebox{\resizebox{0.20\textwidth}{!}{\input{pics/singularity0.pstex_t}}}                            %%% singularities 
\caption{{\em links} and {\em hyperlinks}}\label{fig:links}
\end{center}
\end{figure}

Graphically links are represented like in the left hand side picture of Figure~\ref{fig:links} and distinguished by their shape: triangular for negative and round for positive links. 
We use variables $x^p,y^p,z^p,...$ with polarity  $p\in\{+,-\}$ for links. 

%%%%%%%%%%%%%%%%%%%%%%%%%%%%%%%%%%%%%%%%%%%%%%%%%%

Intuitively, negative links correspond to generalized ($n$-ary) $\parsym$-links while positive links correspond to generalized  $\otimes$-links.
%% Bottom places are labelled by occurrences of positive atoms, while top places are labelled by occurrences of negative atoms. 
In order to capture the additive behavior (a non-deterministic "sharing nature") we need to allow superposition of links; this will naturally bring us to the next notion 
of hyperlinks. 
%% In order to distinguish each multiplicative component ({\em slice}) of an hyperlink we need to equip each hyperlink with {\em weights}.

%%% \begin{definition}[weight]\label{def:weights}
%Assume a set of Boolean variables denoted by $p,\bar p,q,\bar q,...$, then a {\em (monomial) weight} $v, w...$ is a product ``.'' (conjunction)
% of variables or negation of variables. As notation, we use:
%$\epsilon_p$ for a variable $p$  or its negation $\bar p$, $1$ for the empty product, 
%$0$ for a product where both $p$ and $\bar p$ appear.  Two weights, $v$ and $v'$, are {\em disjoint} when $v.v' = 0$; 
%a weight $v$ {\em depends on a variable} $p$ when $\epsilon_p$ appears in $v$. A {\em valuation} 
%$\varphi$ is a function that assigns to each variable one among the two values $\{0,1\}$.
%%% \end{definition}

\begin{definition}[hyperlinks]\label{def:hyperlinks}
 An  {\em hyperlink} is a set of  links that share some (at least one) places as follows:
% each link $x_{i=1,...,n}$ is equipped with a monomial weight $w(x_i)$; 
%the weights associated to the links 
%%% $\lambda_i,\lambda_j$ 
%must be pairwise disjoint, that is $\forall i,j, 1\leq i,j\leq n, w(x_i).w(x_j)=0$;
% a place $l$ takes the same weight of the link whose it belongs to, 
%except when $l$ is shared by a set of links $x_1,...,x_m\subseteq\lambda$, in this case  
%the weight of $l$ is  $w(l)=\sum^{j=1}_m w(x_j)$, moreover, it must be monomial.
\begin{itemize}
\item a {\em negative hyperlink} contains  only negative links and an unique bottom place; %% (all its negative links share this unique bottom place); 
all its jump places must be distinguished (i.e., its negative links have no  jump places in common).

%\begin{enumerate}
% \item it has {\em exactly one}  bottom place (all negative links share the same bottom place);
% \item it is equipped with a set of {\em eigen weight variables} $p_1,...,p_m$ (new for each negative hyperlink);
% \item if $v$ is the weight of the bottom place of $\lambda$, then each negative link $x_{i=1,...,n}$ has weight 
%       $v.w(x_i)$, where $w(x_i)$ is an element (monomial) of the Boolean algebras generated over 
%%% the set of eigen weight variables 
%$p_1,...,p_m$.
%\item for all eigen weight variable $p_i$, the sum of all weights in $\lambda$ depending on $p_i$ must be a monomial.
%\end{enumerate}
\item a {\em positive hyperlink} contains only positive links and  {\em at least one}  bottom place; all its jump places must be distinguished (i.e., its positive links have no  jump places in common).
\end{itemize}
\end{definition}

\begin{figure}
\begin{center}
\resizebox{0.25\textwidth}{!}{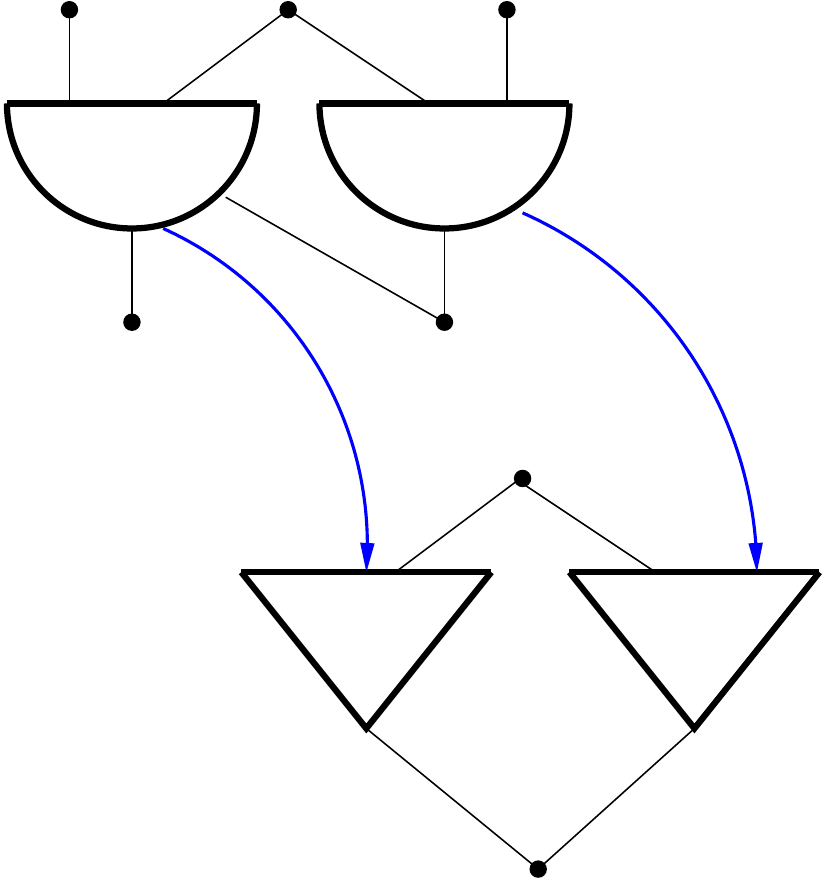}
\hspace{2cm}
\resizebox{0.20\textwidth}{!}{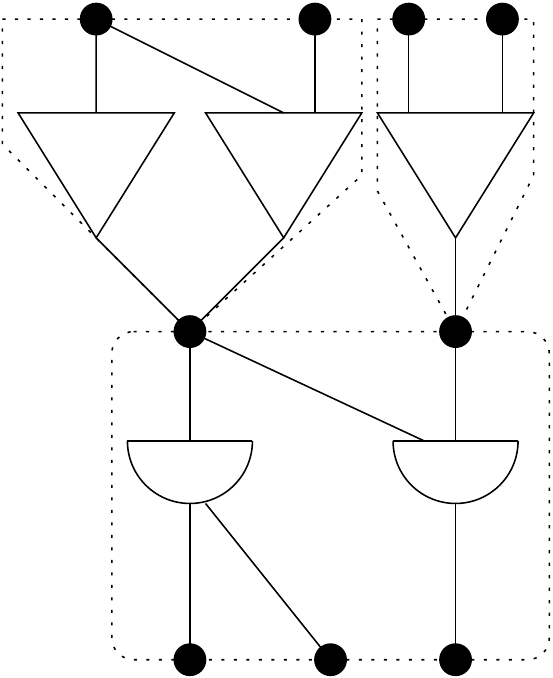} 
\hspace{2cm}
\resizebox{0.20\textwidth}{!}{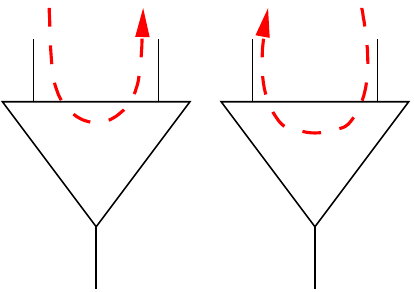} 
\caption{{\em toggling} (left side), {\em bipole} (middle side) and {\em singularities} (right side)}\label{fig:bps}
\end{center}
\end{figure}

Analogously to (multiplicative) links,  negative hyperlinks correspond to generalized  $\&$-links (additive conjunction) while positive links correspond to generalized   $\oplus$-links (additive sum). 
Recall that in linear logic the additive connectives capture non deterministic computational phenomena (typically of distributed middleware systems).
An example of negative (resp. positive) hyperlink is depicted in the middle (resp., right) hand side of Figure~\ref{fig:links}. Observe that these links represent, graphically, 
the  distributive law of negative $\parsym/\&$ (resp.,  positive $\otimes/\oplus$) connectives.
The notation  $X^+$ (resp., $X^-$) denotes a positive (resp., a negative) hyperlink. Moreover we say that:
\begin{itemize}
\item
an edge is called  a {\em jump edge} (simply {\em jump}) when it goes from a positive jump place to a negative jump place;
\item
a (positive) link $x^+$ {\em depends on} a (negative) link $y^-$ when there exists a jump edge that goes from $x^+$ to $y^-$; 
\item
a pair of positive links $x^+_1$ and $x^+_2$ belonging to a same $+$hyperlink $X^+$ is {\em toggled} by a negative hyperlink $Y^-$, if  there exist two negative 
  links $y^-_1,y^-_2$ in $Y^-$, s.t. there is a jump from $x^+_1$ to $y^-_1$ and a jump  from $x^+_2$ to $y^-_2$. 
  
  A graphical interpretation of the toggling condition with  jump edges is then
  given in the picture on the left hand side of Figure~\ref{fig:bps}.
%\begin{figure}[h]
%\begin{center}
%\resizebox{0.20\textwidth}{!}{\input{pics/toggling.pdf_t}}
%\caption{toggling condition}\label{fig:togglig}
%\end{center}
%\end{figure}
\end{itemize}
%We call {\em positive} (resp., {\em negative}) {\em toggling pair} any couple of positive (resp., negative) links belonging to a same positive (resp., negative) hyperlink. We call {\em jump edge} 
%any edge that connects two jump places. We say that a (positive) link $x$ {\em depends on} a (negative) link $y$ when there exists a jump edge from $x$ to $y$. 
Observe that jumps play here the same role (dependency)  {\em eigen weights} play in~\cite{gir96}.

\begin{definition}[bipolar focussing proof structures]\label{def:ps}
 A MALL focussing proof structure (shortly, BPS) is a set $\pi$ of hyperlinks satisfying the following conditions:
 \begin{enumerate}
  \item
  % {\em (disjunction)} 
  the sets of top (bottom) places of any pair of hyperlinks  are disjoint;
  \item
  % {\em (polarity)} 
  if two hyperlinks are adjacent, then they have opposite polarity;
  %% \item jump places $\star$ may occur only on the top of a negative link with the condition that no more than one jump place may occur on the top of a negative link;
  \item
  % {\em (toggling)} 
 in any $+$hyperlink every pair of  links  is toggled by a $-$hyperlink;
  \item jump places are distinguished (links do not share jump places).
  %  \item\label{mon-cond} for each $\epsilon_p$ occurring in the weight $w$ of a positive link $\lambda$, there exists a negative link $\lambda'$ 
%       %% (belonging to some negative hyperlink $\Lambda'$) 
%       s.t. $w(\lambda)\leq w(\lambda')$; we say that $\lambda$ {\em depends (w.r.t. $\epsilon_p$) on} $\lambda'$.
%       %% (that is the analogous of the {\em dependence or monomiality condition} of~\cite{gir96}).
%  \item the bottom places of $\pi$ must have weight $1$.
 \end{enumerate}
 %%\end{definition}
%A proof structure is said {\em bipolar} if any place occurring at the top of some positive hyperlink of $\pi$ also occurs at the bottom 
%of some negative hyperlink of $\pi$ and vice-versa (the place occurring at the bottom of any negative hyperlink, also occurs at the top of 
%some positive hyperlink). 
 \end{definition}

% \begin{description}
%  \item{\em (disjunction)} the sets of top (resp., bottom) places of any two hyperlinks of $\pi$ are disjoint;
%  \item{\em (polarity)} if two hyperlinks are adjacent, then they have opposite polarity;
%  %% \item jump places $\star$ may occur only on the top of a negative link with the condition that no more than one jump place may occur on the top of a negative link;
%  \item{\em (toggling)} every toggling pair of positive links must be separated by a negative hyperlink, i.e., if $x^+_1$ and $x^+_2$ are two toggling links, then there must exist two toggling negative 
%  links $y^-_1,y^-_2$ s.t. there is a jump edge from $x^+_1$ to $y^-_1$ and a jump edge from $x^+_2$ to $y^-_2$. The only jumps are those ones coming from toggling condition. %  \item\label{mon-cond} for each $\epsilon_p$ occurring in the weight $w$ of a positive link $\lambda$, there exists a negative link $\lambda'$ 
%%       %% (belonging to some negative hyperlink $\Lambda'$) 
%%       s.t. $w(\lambda)\leq w(\lambda')$; we say that $\lambda$ {\em depends (w.r.t. $\epsilon_p$) on} $\lambda'$.
%%       %% (that is the analogous of the {\em dependence or monomiality condition} of~\cite{gir96}).
%%  \item the bottom places of $\pi$ must have weight $1$.
% \end{description}

%A proof structure is said {\em bipolar} if any place occurring at the top of some positive hyperlink of $\pi$ also occurs at the bottom 
%of some negative hyperlink of $\pi$ and vice-versa (the place occurring at the bottom of any negative hyperlink, also occurs at the top of 
%some positive hyperlink). 
Finally, $\pi$ is said to be {\em elementary} if it is bipolar and contains exactly a positive hyperlink: each elementary focussing proof structure 
corresponds to a bipole (see the picture in the middle side of Figure~\ref{fig:bps}).

%Observe that condition~\ref{mon-cond} of the Definition~\ref{def:ps} is the analogous of the {\em monomiality condition} of~\cite{gir96}; moreover, an elementary focussing proof structure corresponds to a bipole; finally, a pure {\em multiplicative proof structure} is a proof structure built only on (positive and negative) links whose weight is $1$ 
%(i.e., all hyperlinks are reduced to simple links).

%\begin{example}%%[elementary bipolar proof structure]
%\label{ex:bipolar}
%An example of an elementary (bipolar)  proof structure is given in the middle side picture of %%Figure~\ref{fig:bipolar};
%Figure~\ref{fig:links}; 
%observe the negative layer  contains  two negative hyperlinks (inside dotted lines): that one on the left hand side is equipped with an unique eigen weight variable $r$.
%%\begin{figure}%%%%%%%%%%%%%%%%%%%%%%%%%%%%%%%%%%%%%%%%%%%%%%%
%%\begin{center}
%%\resizebox{0.20\textwidth}{!}{\input{pics/bipolar00.pstex_t}}  
%%\caption{an elementary (bipolar) proof structure}\label{fig:bipolar}
%%\end{center}
%%\end{figure}%%%%%%%%%%%%%%%%%%%%%%%%%%%%%%%%%%%%%%%%%%%%%%%%%%
%\end{example}

We are interested on those (correct) proof structures that correspond to bipolar focussing sequent proofs: 
these are called bipolar proof nets. Before introducing these, we need some technical stuff.

A hyperlink  $X$ (or simply, a link) is said to be {\em just below} (resp., {\em just above}) 
an hyperlink $Y$
%%, with notation $\lambda_1\nnearrow\lambda_2$ (resp.,$\lambda_1\sswarrow\lambda_2$),
 if there exists a place that is both at the top (resp., bottom) of $X$ and at the bottom (resp., top) of $Y$. 
Two hyperlinks are said {\em adjacent} if one is just below (resp., just above) the other. Then, fixed a BPS $\pi$:
\begin{itemize}
 \item a {\em $\&$-resolution} is a choice of exactly one negative link for each negative hyperlink (all the other negative links will be erased);
 \item a {\em slice} $S(\pi)$ for $\pi$ is the graph obtained from $\pi$ after the {\em erasing} induced by a $\&$-resolution, as follow:
%% \begin{itemize}
%% \item 
{\em (i)} a place is erased if all the top (bottom) links sharing it are erased; 
%% \item 
{\em (ii)} a link is erased when at least one of its places is erased.
%% \end{itemize}
%% \item a {\color{blue}multiplicative slice} is a slice with only flat  $+$ or $-$ links.
\item a {\em trip} $T$ in a slice $S(\pi)$ for $\pi$ is a non-empty binary relation on $|S|$ 
(the set of link of $S$) which is finite, connected and s.t. any link $x\in |S|$ has at most one successor 
(resp., one predecessor), if it exists. Then, a {\em negative middle link} $x$ (with a predecessor and a successor)
of a {\em proper trip} $T$ (not reduced to a loop with only two links) is a {\em singularity} 
for $T$ iff $T$ {\em enters $x$ downwards} and {\em exists $x$ upwards} (graphically, $T$ bounces on $x$, like in the right hand side picture of Figure~\ref{fig:bps}).
\end{itemize}

\begin{definition}[bipolar focussing proof net]\label{def:pn}
%A MLL bipolar focussing proof proof structure $\pi$ is {\em correct} (i.e., it is a MLL {\em proof net}) iff 
%any proper loop trip contains at least a singularity. 
A BPS  $\pi$  of MALL is {\em correct}, i.e., it is a {\em bipolar proof net} (BPN) iff any proper loop trip in any slice $S(\pi)$ contains at least a singularity. 
\end{definition}

\begin{figure}
\begin{center} 
\resizebox{0.28\textwidth}{!}{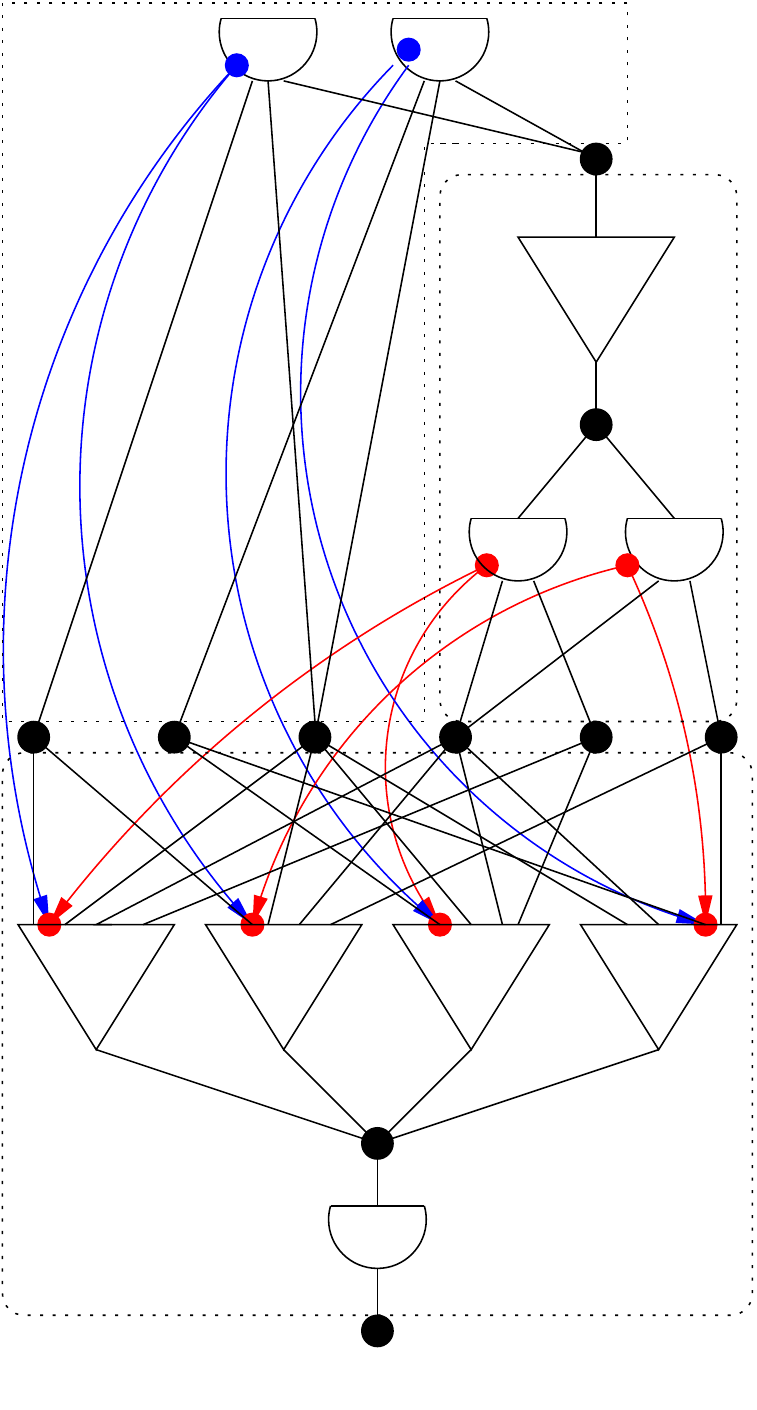}
\hspace{1cm}$\Rightarrow_S$\hspace{1cm}
\resizebox{0.28\textwidth}{!}{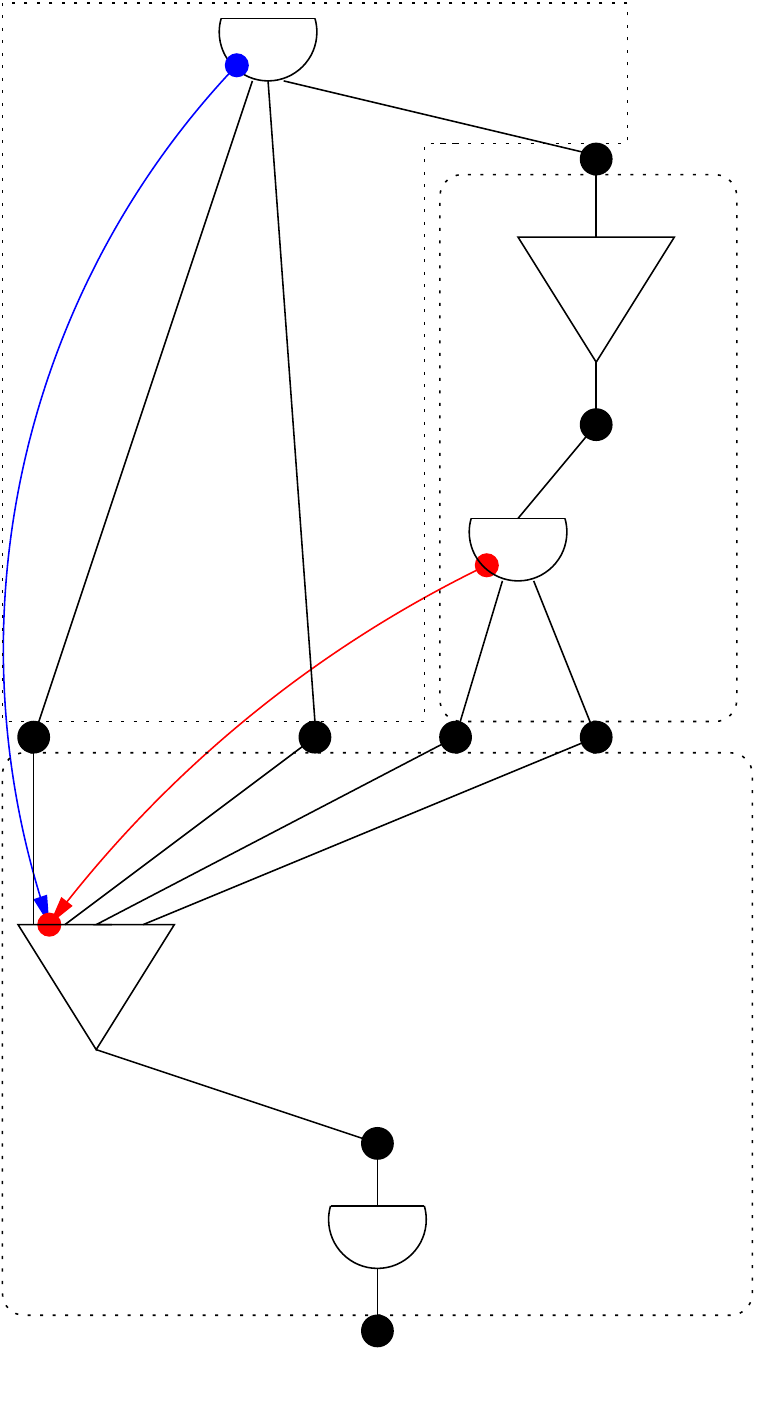}
\caption{a {\em bipolar poof net} (left hand side) with a {\em slice} (right hand side)}\label{fig:bpn}
\end{center}
\end{figure}

An instance of BPN is given in the left hand side picture of Figure~\ref{fig:bpn}. It is not difficult to check that any proper loop trip in any slice contains at least a singularity, in particular that is 
true for the slice depicted in the right hand side of Figure~\ref{fig:bpn}. In order to simplify the reading of these pictures, jumps from positive to negative links are drawn as oriented (colored) curved edges.

We can set a precise correspondence between sequent proofs and proof nets: in the literature this correspondence is called  "sequentialization".

%\subsection{(De-)Sequentialization}
%%%%%%%%%%%%%%%%%%%%%%%%%%%%%%%%%%%
%\label{ssec:deseq}

\begin{theorem}[(de-)sequentialization]\label{th:de-seq}
 A bipolar focussing sequent proof $\Pi$ of $\Gamma$ can be {\em de-sequentialized} in a canonical way 
into a bipolar focussing proof net $\pi$ with same conclusion $\Gamma$ and vice versa, a proof net $\pi$ can be 
{\em sequentialized} into a sequent proof $\Pi$ with same conclusions.
\end{theorem}
The de-sequentialization part of Theorem~\ref{th:de-seq} is proved by induction on the size of the given sequent proof (i.e. the number of  bipoles). For the base of the induction, 
there exists precise correspondence between a bipole and an elementary proof proof structure which is trivially correct (i.e. a proof net). 
As an instance, observe the focussing bipolar sequent proof of $\eta_F$ of the Example~\ref{example:2} de-sequentializes into the bipolar proof net drawn in the left hand side picture of Figure~\ref{fig:bpn} and vice versa. Actually, in order for the bipole $\nu(H)$ to correspond to an elementary focussing proof structure, there is need to introduce a dummy 
negative link with one top place for $c$. This could have been avoided by explicitly introducing a {\em polarity inverter}, as usually done 
in strictly polarized syntax (see~\cite{lau99}). 

The sequentialization part is proved by induction on the number of slices of  $\pi$; observe that a BPN reduced to a single slice is trivially a MLL BPN which 
can be shown that sequentializes into a MLL sequential proof. The crucial task is to show how to gluing the multiplicative (MLL) sequential slices into an additive (MALL) sequential proof.

%%In the following, except when differently mentioned, the term proof net (resp., proof structure) means BPN (resp., BPS).

%\begin{figure}
%\begin{center}
%\fbox{\resizebox{0.32\textwidth}{!}{\input{pics/bipolar0010.pstex_t}}}\hspace{1cm}
%\fbox{\resizebox{0.44\textwidth}{!}{\input{pics/isola002.pstex_t}}}
%%% \fbox{\resizebox{0.32\textwidth}{!}{\input{pics/bip-exp00.pstex_t}}}
%\end{center}
%\caption{a {\em bipolar focussing proof net} (left hand side) and the {\em isolation property} (right hand side)}\label{fig:bipolaris}
%\end{figure}

\smallskip
In the next section we study the problem of constructing a proof net by a juxtaposition of  concurrent bipoles (agents).
This proof net construction can be viewed as a computational paradigm for middleware (infrastructure) programming.

\section{Proof Net Construction as a Middleware Paradigm}
%%%%%%%%%%%%%%%%%%%%%%%%%%%%%%%%%%%%%%%%%%%%%%%%%%%%%%%%%%%%%%%%%
\label{sec:bpncmp}

%% Bipoles always drive the construction bottom-up by hyperlink expansion and it is driven by ''available`` top places of the structure.
In  building a proof net, places (except $\star$) are decorated by type informations (occurrences of negative atoms); each bipole 
is viewed as a disjoint sum of collaborative agents which continuously attempt  to perform a construction step, that is, an 
expansion of the proof net obtained by adding  an elementary bipolar proof structure (a bipole) from the places whose types match the trigger of the given bipole.  
Bipoles always drive the construction bottom-up like in the left hand picture of  %%Figure~\ref{fig:expstep}
Figure~\ref{fig:case01}.
%\begin{figure}
%\begin{center}
%\resizebox{0.32\textwidth}{!}{\input{pics/bip-exp00.pstex_t}}
%\caption{an expansion step}\label{fig:expstep}
%\end{center}
%\end{figure}

An expansion step is  correct if it preserves the property of 
being a proof net. Checking correctness (singularity-free trips)  is a task which may involve 
{\em visiting a large portion} of the expanded proof structure.
Now, since this construction is performed collaboratively and concurrently by a cluster of bipoles
for true concurrency we need to:\\
%% \begin{enumerate}
%% \item\label{isola1} 
1) restrict the traveling region (reducing possible conflicts among agents);\\
%% \item\label{isola2} 
2) protect (lock) the gathered information against attempts of other concurrent agents;\\
%% \item\label{isola3} 
3) increment/update, in case of success, the locked information for transition.
%% \end{enumerate}

Good bounds for these tasks are necessary; however, in the following two sub-sections~\ref{ssec:maximal} and~\ref{ssec:domination} we mainly focus on the task~1. 

\subsection{Maximal switchings}
%%%%%%%%%%%%%%%%%%%%%%%%%%%%%%%%%%%%%%%
\label{ssec:maximal}

First we show that, in order to detect  singularity-free trips we may restrict us to consider only particular subgraphs  of switchings, 
these are called maximal switchings.

\begin{definition}[maximal switching]\label{def:top-switch}
A jump edge from $x^+$ to $y^-$ is said {\em maximal} in a switching $S(\pi)$ if there not exists in such a switching 
a positive link $z^+$ such that it depends on $y^-$ too and it is above $x^+$ in $\pi$; then a {\em maximal switching} 
is a switching containing only maximal jump edges.
\end{definition}

\begin{lemma}[maximal switchings]\label{lem:eff-corr}
A proof structure $\pi$ is correct 
%% (according to Definition~\ref{def:pn})
 iff any proper loop of any maximal switching $S$ for $\pi$ contains at least a singularity. 
\end{lemma}

\begin{figure}
\begin{center}
\fbox{\resizebox{0.40\textwidth}{!}{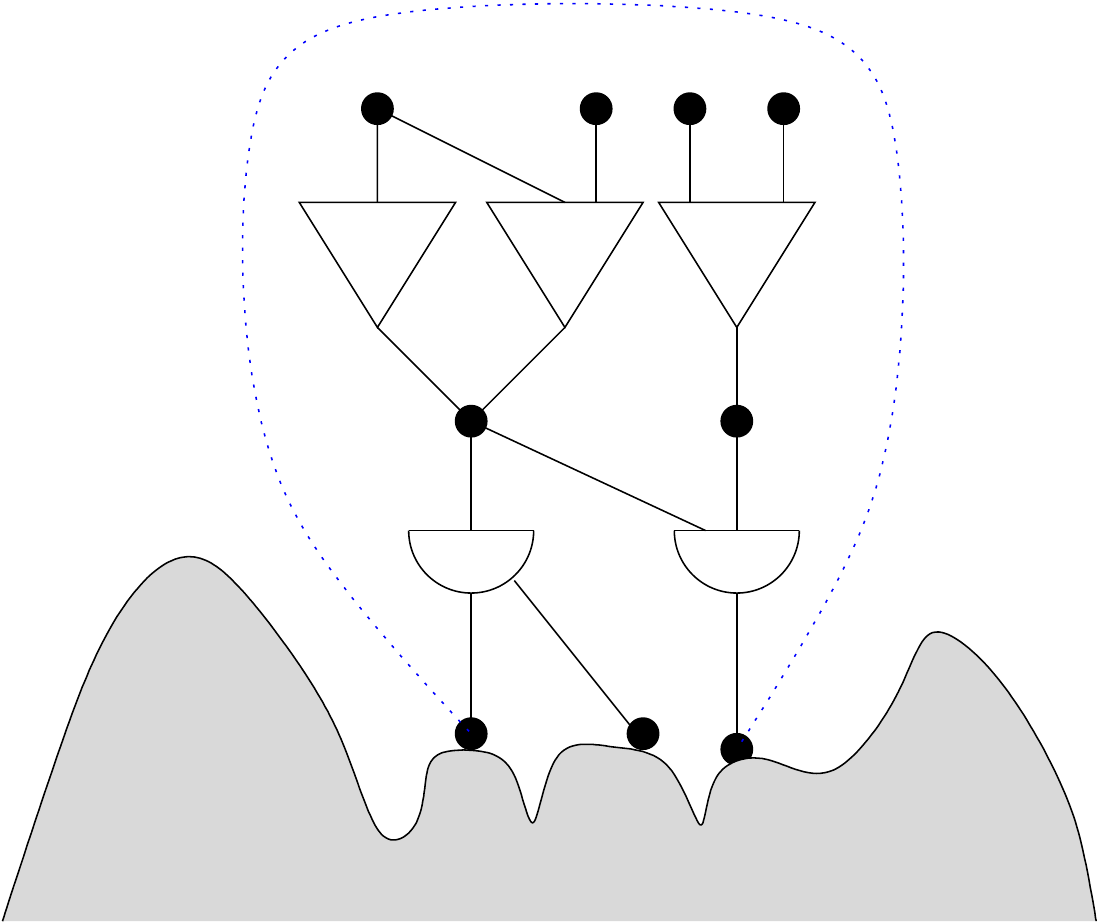}}
\hspace{1cm}
\fbox{
\resizebox{0.18\textwidth}{!}{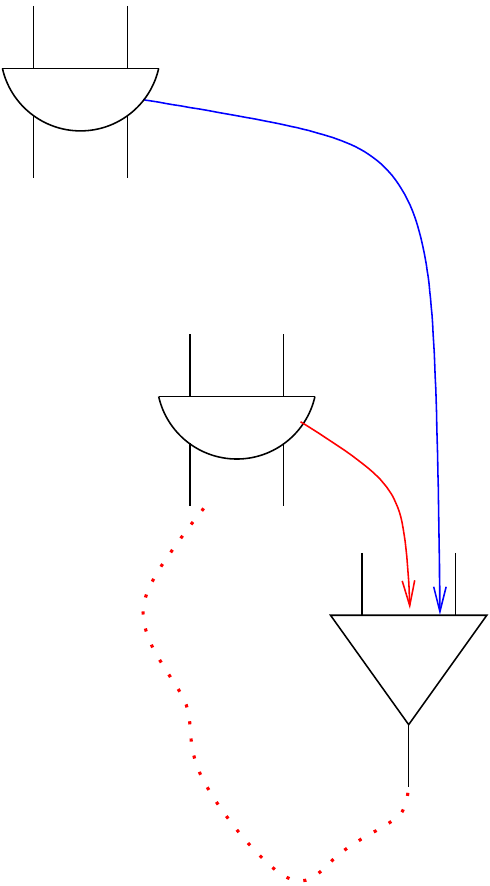}
\hspace{0.1cm}
$\Rightarrow$
\hspace{0.1cm}
\resizebox{0.22\textwidth}{!}{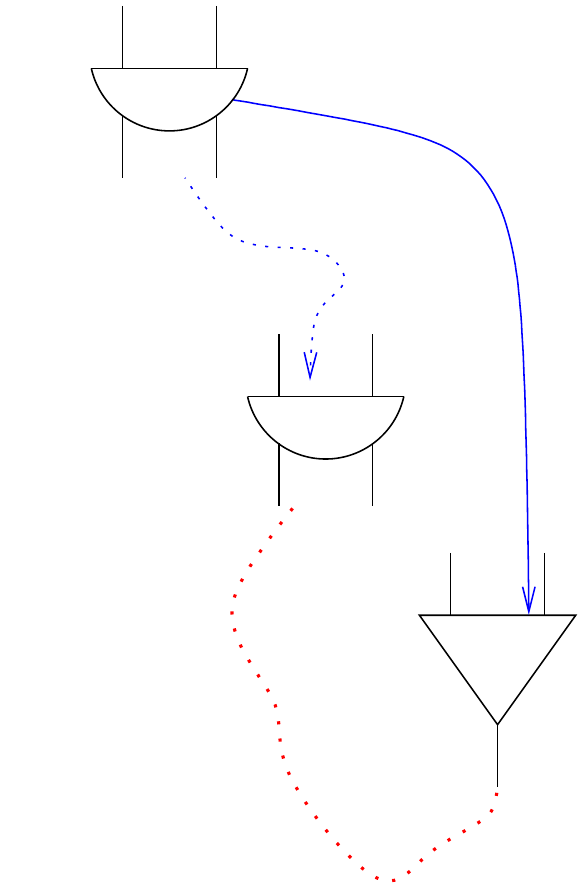}
}
%\hspace{0.1cm}
%\fbox{\resizebox{0.46\textwidth}{!}{\input{pics/isola002.pstex_t}}}
\end{center}
\caption{{\em expansion step} (left hand side) and {\em maximal switching} (right hand side)}\label{fig:case01}
\end{figure}

By Definition~\ref{def:top-switch}, if  there exist in $S(\pi)$ two positive links,  $x^+$ and $z^+$, both depending 
on $y^-$ and with $z^+$ above $x^+$, then there must exist in $S(\pi)$ a path going 
from $x^+$ upwards to $z^+$; clearly, if there exists a singularity-free loop $T$ in $S(\pi)$ containing a jump from $x^+$ to $y^-$, there will also exist 
a singularity-free loop $T'$ in $S(\pi)$ containing a jump from $z^+$ to $y^-$ (see the right hand side of picture of Figure~\ref{fig:case01}).

\subsection{Expansion under domination}
%%%%%%%%%%%%%%%%%%%%%%%%%%%%%%%%%%%%%%%
\label{ssec:domination}

We fix once for all a maximal switching $S$ for $\pi$,  then we show (Lemma~\ref{lem:isola})  that, w.r.t. a candidate expansion, only certain negative links must be 
explored (isolated and locked); the other ones are available for other possible expansions (or transactions). 

\begin{definition}[domination order]\label{def:domin}
Assume $x,y$ are two negative links in a  switching $S$ for $\pi$; 
a {\em root} of $S$ is any (positive) link of $S$ that has no link 
below it. Then, $x\leq y$ ({\em $x$ dominates $y$}) 
if any singularity-free trip starting at a root and stopping 
upwards at $y$ visits $x$ upwards.
\end{definition}

\begin{proposition}[forest order]\label{prop:fo}
The relation $\leq$ on negative links of $S$ is a {\em forest order}; 
it is reflexive, antisymmetric, transitive and 
it satisfies the following property on negative links:
$$\forall x,y,z\;\;if\;\;(x\leq z\;\wedge\;y\leq z)\;\;then\;\;(x\leq y\;\vee\;y\leq x).$$
\end{proposition}

The  {\em joint dominator of $N$}, $\bigwedge(N)$, 
is the {\em greatest lower bound} (g.l.b., when it exists), 
by $\leq$, of a set of negative links $N$.

If the set of the predecessor by $<$ of a negative link $x$ is not empty, 
then it has a greatest element, by $\leq$, called the {\em  immediate dominator} $D(x)$ of a negative link $x$.

\begin{lemma}[isolation property]\label{lem:isola}
%% {\sc Theorem (lower bound):} 
Let $x,y$ be two negative links and $T$ be 
a singularity-free trip of a  switching $S$ for $\pi$ starting 
downwards at $x$ and stopping upwards at $y$; 
then, any negative link $z$ visited by $T$ is {\em strictly dominated} 
by the joint dominator $\bigwedge\{\lambda_1,\lambda_2\}$ (if defined):
$$\forall z^-\in |T|,\;\bigwedge\{x,y\} < z.$$
\end{lemma}

\begin{figure}
\begin{center}
\resizebox{0.45\textwidth}{!}{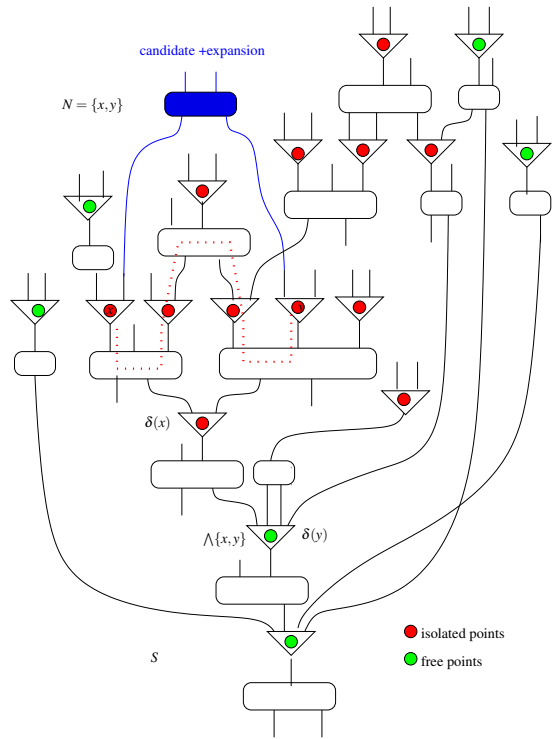}
\caption{proof net interpretation of the {\em isolation property}}\label{fig:isola}
\end{center}
\end{figure}

Clearly, w.r.t. an expansion of a proof net $\pi$ by a bipole $\beta$, Lemma~\ref{lem:isola}  gives a good (lower) bound to the region to be explored in order to detect 
a singularity-free trip in a  switching $S(\pi*\beta)$ (where $\pi*\beta$ denotes the juxtaposition of $\beta$ over $\pi$). 
An instance of a candidate multiplicative expansion that is not correct is given in Figure~\ref{fig:isola}: grey (or red) negative links denote all 
those negative links that (according to Lemma~\ref{lem:isola}) must be visited in order to look for a singularity-free loop inside 
a top switching; while the light grey (or green) ones are unexplored and so available for other transactions.

We propose in the next (last) subsection some applications of Lemmas~\ref{lem:eff-corr}  and~\ref{lem:isola}  to the theoretical interpretation of  distributed transactional systems.

\subsection{Transactional Systems}
%%%%%%%%%%%%%%%%%%%%%%%%%%%%%%%%
\label{ssec:transactions}

A transaction combines a group of independent actions into a single action with a set of predictable outcomes. 
Traditionally, transactions are required to adhere to the {\em ACID} properties of {\em Atomicity} (ensuring that all
actions in the transaction either complete successfully, or revert to a state where none of them were                            
run), {\em Consistency} (ensuring that the system is not put into an illegal state), {\em Isolation} 
(letting concurrent transactions run as if they were the only transaction being processed), and {\em Durability} (ensuring that
any completed transaction has its stable outcome and cannot be undone, even by accidental hardware or software failure). 
However, while transaction management in traditional systems typically offers an acceptable level of service, the same cannot be said for transactions
achieved by combining services offered by multiple systems. Such multi-databases transactions often run for much longer periods of time than traditional 
transactions, so locking any data may block other transactions for an unacceptable length of time.
Because of this, the traditional ACID properties are typically reduced in strength, helping to ensure that 
the entire system maintains an acceptable level of service. 
Typically, in the Web Services environment, traditional ACID transactions are not always sufficient to support the activities that businesses would like to process. 
Transactions that involve multiple service providers can run for long periods of time. This can result in negative side-effects when combined with traditional
transaction-based concurrency control mechanisms. While Web Services transactions standards do exist, it is still difficult 
(e.g., for an end-user) to combine services from loosely-coupled providers so that they are used as a single {\em co-operative transaction} (\cite{webster08}). 

Under this respect, the paradigm of proof net construction can be put in correspondence  with transactional systems paradigms. 
That can be seen as an analogous of the well known {\em Curry-Howard correspondence} between the cut-reduction paradigm and the functional programming paradigm.
Any correct expansion step can be seen as a transaction; more precisely: 
\begin{itemize}
 \item %% transaction in pure MLL proof nets correspond to {\em ACID transactions} and in particular
       Lemma~\ref{lem:isola} captures the {\em multiplicative behavior} of proof nets and corresponds to the {\em isolation property} of ACID transactions. 
       For instance, Figure~\ref{fig:isola} can be interpreted as  a candidate multiplicative expansion that is  not an ACID transaction (it is not correct);
\item Lemmas~\ref{lem:eff-corr} captures the {\em additive behavior} of proof nets and corresponds to {\em co-operative transactions}: actually, 
we can additively  "slice"  a transaction in to a sum of interacting (or cooperative) ACID transactions; maximality of switching guarantees that 
only certain resources will be locked.
\end{itemize}

%%%%%%%%%%%%%%%%%%%%%%%%%%%%%%%%%%%%%%%%%%%%%%%%
%%%%%%%%%%%%%%%%%%%%%%%%%%%%%%%%%%%%%%%%%%%%%%%%
\bibliographystyle{plain} 

\begin{thebibliography}{1}

\bibitem{jma01}
J.-M. Andreoli.
\newblock Focussing and Proof Construction.
\newblock {\sl Annals of Pure and Applied Logic} 107(1), pp 131--163, 2001.

\bibitem{jma02}
J.-M. Andreoli.
\newblock Focussing proof-nets construction as a middleware paradigm.
\newblock In A. Voronkov, ed., {\sl Proc. of the 18th Int'l Conference on Automated Deduction. Lecture Notes 
in Computer Science, pp. 501-516}. Denmark, 2002. Springer Verlag.

\bibitem{jma03}
J.-M. Andreoli and L. Mazar\`e.
\newblock Concurrent Construction of Proof-Nets.
\newblock {\sl In proc. of Computer Science Logic (CSL)}, Wien, Austria, 2003.

\bibitem{gir87}
J.-Y. Girard. 
\newblock Linear Logic.
\newblock {\sl Theoretical Computer Science}, 50:1--102, 1987.

\bibitem{gir96}
J.-Y. Girard. 
\newblock Proof-nets: the parallel syntax for proof theory. 
\newblock {\sl Logic and Algebra}, Marcel Dekker, 1996.

\bibitem{gir01}
J.-Y. Girard. 
\newblock Locus Solum. 
\newblock {\sl Mathematical Structures in Computer Science} 11, pp. 301-506, 2001.

\bibitem{lau99}
Laurent, O. \
{\em Polarized Proof-Nets: Proof-Nets for LC (Extended Abstract)}.
 In J.-Y. Girard, editor, Typed Lambda Calculi and Applications 1999, 
LNCS 1581, pp. 213-227. Springer-Verlag. Avril 1999.

\bibitem{miller96}
D. Miller.
\newblock Forum: a multiple-conclusion specification logic.
\newblock {\sl Theoretical Computer Science} 165, pp. 201-232, 1996.

\bibitem{webster08}
D. Paul, M. Wallis, F. Henskens and M. Hannaford.
\newblock Transaction support for interactive web applications.
\newblock {\sl Proceedings of the 4th International Conference on Web Information Systems and Technologies (WEBIST 2008)}. 
Funchal-Madeira, Portugal 4-7 May, 2008.

\end{thebibliography}

\end{document}